\documentclass[a4paper,12pt]{article}

\usepackage{amsmath}\usepackage{amssymb}
\usepackage{latexsym}\usepackage{cite}

\newcommand{\be}{\begin{equation}}
\newcommand{\ee}{\end{equation}}

\def\beq{\begin{equation}}
\def\eeq{\end{equation}}

\def\al{\alpha}

\def\de{\delta}
\def\De{\Delta}

\def\te{\theta}

\def\om{\omega}
\def\ep{\epsilon}

\def\sq{\sqrt}

\def\l{\left (}
\def\r{\right )}

\def\fr{\frac}
\def\la{\label}
\def\hs{\hspace}
\def\vs{\vspace}

\def\ran{\rangle}
\def\lan{\langle}

\def\tl{\tilde}
\def\tm{\times}

\begin{document}

\begin{flushright}
OSU-HEP-07-04\\
August 29, 2006 \\
\end{flushright}

\vs{0.5cm}

\begin{center}
{\Large\bf

New Prediction For Leptonic $\te_{13}$
}
\end{center}

\vspace{0.5cm}
\begin{center}
{\large
{}~S.~Nandi{}~ and ~Zurab Tavartkiladze
}
\vspace{0.5cm}

{\em Department of Physics and Oklahoma Center for High Energy
Physics, Oklahoma State University, Stillwater, OK 74078, USA }

\end{center}
\vspace{0.6cm}

\begin{abstract}


An extension of the neutrino sector with two right handed singlet
neutrinos responsible for Dirac neutrino masses is discussed. We
show that this setup with flavor symmetry can give large solar and
atmospheric neutrino mixings and suppressed $\te_{13}$. The flavor
symmetry ${\cal U}(1)\tm S_4$ is shown to lead to $\te_{23}\simeq
\pi/4$ and a new predictive formula for the $\te_{13}(\simeq
0.015)$.

\end{abstract}

\vs{0.7cm}


\newpage

Recent atmospheric \cite{Fukuda:2000np} and solar
\cite{Fukuda:2001nj} neutrino data have confirmed the  neutrino
oscillations. A global analysis \cite{Maltoni:2004ei}
for the oscillation parameters yields
$$
|\De m_{\rm atm}^2|=2.4 \cdot \l 1^{+0.21}_{-0.26}\r \tm 10^{-3}~{\rm eV}^2~,~~~
\sin^2\te_{23}=0.44 \cdot \l 1^{+0.41}_{-0.22}\r ~,
$$
\beq
\De m_{\rm sol}^2=7.92 \cdot \l 1\pm 0.09\r \tm 10^{-5}~{\rm eV}^2~,~~~
\sin^2\te_{12}=0.314 \cdot \l 1^{+0.18}_{-0.15}\r ~.
\la{atm-sol-data}
\eeq
The third leptonic mixing angle
$\te_{13}$ has not yet been measured  but a useful upper bound exists,
\beq
\te_{13}\stackrel{<}{_\sim }0.2~,
  \la{chooz}
  \eeq
provided by the CHOOZ experiment \cite{Apollonio:2002gd}. The
corresponding mass scales in (\ref{atm-sol-data}) indicate new
physics beyond the SM or beyond its minimal SUSY extension - MSSM.
Looking at (\ref{atm-sol-data}) and (\ref{chooz}), together with
neutrino mass generation,  one can try to understand the origin of
two large mixing angles ($\te_{12}\simeq 35^{\rm o}$,
$\te_{23}\simeq 42^{\rm o}$) and a suppressed third angle
($\te_{13}\stackrel{<}{_\sim }12^{\rm o}$). Non-zero value of
$\te_{13}$ would cause the CP violation in the neutrino
oscillations. In addition to the values of $\te_{13}$ and the phase
$\de$, the sign of $\De m_{\rm atm}^2$ is also unknown. This sign is
directly related to whether neutrinos have normal or inverted
hierarchical mass pattern. Planned experiments are expected to shed
more light to these important issues. These will give  new selection
rules for theoretical model building and will rule out many existing
scenarios. On the other hand, it still remains a great challenge to
build self consistent scenarios which give natural explanation of
bi-large neutrino mixings and predict the value of $\te_{13}.$ In
this respect, the symmetry principle seems to be the most powerful
tool and we will pursue this approach here. Numerous attempts have
been made \cite{Shafi:2004jy}-\cite{Babu:2007zm} to explain and/or
predict the suppressed value of $\te_{13}$ within different setups.
In this paper we suggest new framework which naturally gives
bi-large neutrino mixings and suppressed $\te_{13}$. Our proposal
works for both non SUSY and SUSY scenarios. Since supersymmetry has
strong theoretical and phenomenological motivations, we stick here
with the SUSY description. We consider extension of the MSSM by two
right handed neutrino (RHN) superfields $N_{1,2}$ and flavor
symmetry $G_{f}$ (for earlier work with similar extension see \cite{King:1999mb}). An extension of the SM (plus a $Z_2$ symmetry)
with RHNs and an additional Higgs doublet with a tiny VEV  has
recently been considered \cite{Gabriel:2006ns} for understanding the tiny neutrino masses
as an alternative to the usual sea-saw mechanism. The neutrinos are
Dirac particle, and the model of \cite{Gabriel:2006ns} has interesting phenomenological
implications, specially  for the Higgs boson searches.
In this work, our focus is different and thus we restrict ourself
without additional Higgs doublets. Neutrino mass suppression will happen due to
tiny Yukawa couplings($\stackrel{<}{_\sim }10^{-11}$) guaranteed by flavor symmetry.
We show that with $G_f={\cal U}(1)$ the value of $\te_{13}$
is estimated to be $\sim \fr{\De m_{\rm sol}^2}{\De m_{\rm atm}^2}\approx 0.03$, while with $G_f={\cal U}(1)\tm
S_4$, an accurate prediction for $\te_{13}$ is possible
$\te_{13}\simeq 0.014$.

\vs{0.2cm}

{\bf $\bullet $ Case with ${\cal U}(1)$ flavor symmetry}

\vs{0.2cm}

Let us start the discussion with neutrino sector augmented by two
RHNs $N_1, N_2$ and $G_f={\cal U}(1)$ flavor symmetry. For ${\cal
U}(1)$ symmetry breaking we introduce MSSM singlet superfield $X$
charged under ${\cal U}(1)$ which has VEV in its scalar component
\beq \fr{\lan X\ran }{M_{\rm Pl}}\equiv \ep ~. \la{XVEV} \eeq This
is realized naturally if ${\cal U}(1)$ is an anomalous gauge
symmetry often emerging from superstrings \cite{Witten:1981nf}.
Then, $X$'s VEV can be fixed by cancelation condition of the
Fayet-Iliopoulos $D_A$ term: $\xi +Q_X|X|^2=0$.

Consider the following ${\cal U}(1)$ charge assignment:
\beq
Q_X=-1~, ~~ Q_{h_u}=0~, ~~ Q_{l_i}=q~,~~ Q_{N_1}=n+1-q ~,~~Q_{N_2}=n-q~,
\la{U1charges}
\eeq
where $n$ is a positive integer and  $l_i$ denote $SU(2)_L$ lepton doublet superfields, while $h_u$ is up type
MSSM Higgs superfield. The relevant couplings allowed by ${\cal U}(1)$ symmetry and leading to the Dirac neutrino
masses are
\beq
\begin{array}{ccc}
 & {\begin{array}{ccc}
\hs{-0.9cm}N_1 & N_2
\end{array}}\\ \vspace{0mm}
M_{\nu }:=
\begin{array}{c}
l_1\\ l_2\\l_3
\end{array}\!\!\!\!\!\! &{\left(\begin{array}{cc}
\hs{-0.1cm}a\ep ~&0
\\
\hs{-0.2cm}\ep ~ &
\hs{-0.1cm}\al
\\
\hs{-0.1cm} b\ep e^{{\rm i}\de }~&1
\end{array} \right)\ep^n h_u \! }~,
\end{array}
\la{lNhu}
\eeq
where instead of powers of $X/M_{\rm Pl}$ we have substituted $\ep $ according to the notation of
Eq. (\ref{XVEV}) and $a, b, \al $ are dimensionless constants of the order one. We have set (1,2) entry of the matrix in (\ref{lNhu}) to zero. This can be achieved
by proper rotation of $N_{1,2}$ states without any loss of generality, because this basis redefinition
does not change hierarchical structure between remaining matrix elements. Moreover, by proper phase
redefinitions only one complex phase $\de $ remains in the (3,1) entry.

The mass eigenvalues and the unitary matrix transforming the left handed neutrinos upon diagonalization
of (\ref{lNhu}) can be easily found from the diagonalization of the matrix $M_{\nu }M_{\nu }^{\dagger }$. For
convenience we will write it in a form:
$$
\begin{array}{ccc}
 & {\begin{array}{ccc}
 & &
\end{array}}\\ \vspace{2mm}
M_{\nu }M_{\nu }^{\dag }=\begin{array}{c}
\end{array}\!\!\!\!\!\! &{\left(\begin{array}{ccc}
\hs{-0.1cm}0~& 0 ~ &0
\\
\hs{-0.2cm}0~ & \hs{-0.1cm}\al^2~& \al
\\
\hs{-0.1cm} 0~&\al  ~& 1
\end{array} \right)\fr{m^2}{1+\al^2}\! }
\end{array}
\begin{array}{ccc}
 & {\begin{array}{ccc}
 & &
\end{array}}\\ \vspace{1mm}
+~\begin{array}{c}
\end{array}\!\!\!\!\!\! &{\left(\begin{array}{ccc}
\hs{-0.1cm}a^2~& a ~ &abe^{-{\rm i}\de }
\\
\hs{-0.2cm}a~ & 1~& be^{-{\rm i}\de }
\\
\hs{-0.1cm}abe^{{\rm i}\de } ~& be^{{\rm i}\de }~& b^2
\end{array} \right)\fr{m^2\ep^2}{1+\al^2}\! }~,
\end{array}
$$
\beq {\rm with}~~~~~~~~~m^2=(1+\al^2)\ep^{2n}\lan h_u^{(0)}\ran^2~.
\la{MnuMnu}
\eeq
We will take $\ep \ll 1$ which is natural from the symmetry viewpoint and leads to the neutrino mass pattern with
a normal hierarchy.
In (\ref{MnuMnu}) we have split   $M_{\nu }M_{\nu }^{\dag }$ in
two parts. First one including $2\tm 2$  matrix block of rank one
and second one is $3\tm 3$ rank one matrix. The first leading part
is responsible for the mass $m_3$ the heaviest neutrino and for
$\te_{23}$ mixing. The second, sub-leading part, gives $m_2$ and
$\te_{12}$. In particular, we have
\beq
m_1=0~,~~~~
m_2\simeq \sq{\De m_{\rm sol}^2} \simeq \fr{m\ep }{1+\al^2}\left [a^2(1+\al^2)+|1-b\al e^{-{\rm i}\de }|^2\right ]^{1/2} ~,~~~
m_3\simeq \sq{\De m_{\rm
atm}^2}\simeq m~.
\la{masses}
\eeq
Furthermore, for diagonalizing
unitary matrices  we have \beq U_{\nu }^TM_{\nu }M_{\nu }^{\dagger }U_{\nu }^*
=(M_{\nu }^{\rm diag})^2~,~~~~~{\rm
with}~~~~~U_{\nu }=U_{23}U_{13}U_{12}~,
\la{UnTrans}
\eeq
where $U_{\nu }$ is the lepton mixing matrix (assuming that the charged lepton mass matrix is diagonal) and
\beq
\begin{array}{ccc}
 & {\begin{array}{ccc}
 & &
\end{array}}\\ \vspace{2mm}
U_{23}=
\begin{array}{c}
\end{array}\!\!\!\!\!\! &{\left(\begin{array}{ccc}
\hs{-0.1cm}1~,&
0 ~, &0
\\
\hs{-0.2cm}0~, &
\hs{-0.1cm}c_{23}~,& s_{23}
\\
\hs{-0.1cm} 0~,&-s_{23} ~,&
c_{23}
\end{array} \right)\! }~,~~~c_{23}\equiv \cos \te_{23}~, s_{23}\equiv \sin \te_{23}~,~~~\tan \te_{23} =\al ~,
\end{array}
\la{U23}
\eeq
$$
\begin{array}{ccc}
 & {\begin{array}{ccc}
 & &
\end{array}}\\ \vspace{2mm}
U_{12}\simeq
\begin{array}{c}
\end{array}\!\!\!\!\!\! &{\left(\begin{array}{ccc}
\hs{-0.1cm}c_{12}e^{{\rm i}\om }~,& s_{12}e^{{\rm i}\om } ~, &0
\\
\hs{-0.2cm}-s_{12}~, & \hs{-0.1cm}c_{12}~,& 0
\\
\hs{-0.1cm} 0~,&0 ~,&
1
\end{array} \right)\! }~,
\end{array}
~~~
\begin{array}{ccc}
 & {\begin{array}{ccc}
 & &
\end{array}}\\ \vspace{2mm}
U_{13}\simeq
\begin{array}{c}
\end{array}\!\!\!\!\!\! &{\left(\begin{array}{ccc}
\hs{-0.1cm}c_{13}e^{{\rm i}\phi }~,&
0 ~, &s_{13}e^{{\rm i} \phi }
\\
\hs{-0.2cm}0~, &
\hs{-0.1cm}1~,& 0
\\
\hs{-0.1cm}-s_{13}~,&0 ~,&
c_{13}
\end{array} \right)\! }~,
\end{array}
$$
$$
 \phi =-{\rm Arg}\l s_{23}+bc_{23}e^{{\rm i}\de }\r ~,
~~~\om =-\phi -{\rm Arg}\l c_{23}-bs_{23}e^{-{\rm i}\de }\r ~,
$$
\beq \tan \te_{12}\simeq \fr{a}{|c_{23}-bs_{23}e^{-{\rm i}\de
}|}~,~~~~~ \tan \te_{13}~\simeq
\fr{a\ep^2}{1+\al^2}|s_{23}+bc_{23}e^{{\rm i}\de }|~. \la{tans} \eeq
We see that for $a\sim b\sim \al \sim 1$ we have naturally bi-large
mixing $\tan \te_{23}\sim 1$ and $\tan \te_{12}\sim 1$. Moreover,
using (\ref{masses})-(\ref{tans}), we can express the third mixing
angle $\te_{13}$ as
\beq
 \tan \te_{13} \simeq \fr{\De m_{\rm sol}^2}{\De m_{\rm atm}^2} \fr{\tan
\te_{12}\tan \te_{23}}{1+\tan^2 \te_{12}}
\left |  \fr{1+b\cot \te_{23}e^{{\rm i}\de }}{1-b\tan \te_{23}e^{-{\rm i}\de }}\right |  ~.
 \la{13estim}
 \eeq
We see that $\te_{13}$ is suppressed by factor
$\fr{\De m_{\rm sol}^2}{\De m_{\rm atm}^2}~(\ll 1)$, which appears due to the specific texture of Eq. (\ref{MnuMnu}).
The texture similar to
(\ref{MnuMnu}) for active Majorana neutrinos was considered in \cite{te13Texture}, while  in the scenarios of refs. \cite{Chen:2004rr}, \cite{Shafi:2005rd}
was derived by symmetries. Here, however, the neutrinos
are purely Dirac type and the texture (\ref{MnuMnu}) is the
`squire' ($M_{\nu }M_{\nu }^{\dag }$) of Dirac mass matrix $M_{\nu
}$. This cause stronger suppression of the $\te_{13}$ angle
(note however that the model of \cite{Chen:2004rr} deals with Majorana neutrinos and due to $S_3$ flavor symmetry  $\te_{13}$ is still suppressed by
$\fr{\De m_{\rm sol}^2}{\De m_{\rm atm}^2}$ factor).

Although the unknown parameter
$b$ enters in (\ref{13estim}), from the naturalness viewpoint we
expect that $b \sim 1/3 - 3$ and therefore suppression always will
happen giving $\te_{13}\sim 10^{-2}$. The $\te_{13}$  might enhanced
when $|1-b\tan \te_{23}e^{-{\rm i}\de }|\equiv \tl{b}\ll 1$.
However, this  can not be realized because the
 same combination sets the values of $\tan \te_{12}$
 and $\sq{\De m_{\rm sol}^2}$ [see Eqs . (\ref{tans}) and (\ref{masses})].
  The flavor symmetry ${\cal U}(1)$ used here,
besides the suppression of $\te_{13}$, also gives an explanation of
small value of $\fr{\De m_{\rm sol}^2}{\De m_{\rm atm}^2}\sim
\ep^2$, giving $\ep \sim 1/6-1/5$. Also, $\sq{\De m_{\rm
atm}^2}\simeq \ep^n\lan h_u^{(0)}\ran \simeq 0.05$~eV
requires $n=16, 17$. ${\cal U}(1)$ flavor symmetry does not give
hint about values of $b, \de $ and therefore does not allow for
$\te_{13}$'s accurate prediction. In addition, although $\te_{23}$
and $\te_{12}$ are naturally large it is desirable to have more clue (from the theory)
about their values.
  Below we present a more concrete model which gives strict predictions for $\te_{13}$ and $\te_{23}$.

\vs{0.3cm}

{\bf $\bullet $ Case with $G_f={\cal U}(1)\tm S_4$ and prediction of $\te_{13}$.}

\vs{0.2cm}

The expression (\ref{13estim}) obtained above is very intriguing
since it gives naturally suppressed  $\te_{13}$. However, if by some
reason $b=0$ then also $\de $ becomes irrelevant and we will be able
to express $\te_{13}$ in terms of oscillation parameters which are
already measured! We now show that this can be achieved if together
with ${\cal U}(1)$ we introduce non-Abelian discrete symmetry. The
latter have been proven to be powerful in obtaining various
predictive relations \cite{Pakvasa:1977in}, \cite{Barbieri:1999km},
\cite{Babu:2007zm}. We will treat three left handed lepton doublets
$l_i$ as a triplet in a family space. For this purpose either $S_4$
or $A_4$ can be used. For demonstration, we discuss an example with
$G_f={\cal U}(1)\tm S_4$. Thus $\vec{l}\equiv (l_1, l_2, l_3)\sim
{\bf 3_1}$ and $N_{1,2}$ and $h_u$ are $S_4$ singlets ${\bf 1_1}$ \cite{S4ref}.
For $S_4$ breaking and desirable neutrino mass matrix generation we
introduce three scalar superfields $\vec{S}_1, \vec{S}_2,
\vec{A}\sim {\bf 3_1}$. We use the ${\cal U}(1)$ charge assignment
for $X, h_u, \vec{l}, N_1$ as given in (\ref{U1charges}) and take
$Q_{N_2}=n-q-q_A$, $Q_{\vec{A}}=q_A$,
$Q_{\vec{S_1}}=Q_{\vec{S_2}}=0$.
 Then couplings allowed by introduced
symmetries are \beq \fr{\ep^{n+1}}{M_{\rm Pl}}\vec{l}\l
\vec{S}_1+\vec{S}_2\r N_1h_u +\fr{\ep^n}{M_{\rm
Pl}}\vec{l}\vec{A}N_2h_u~. \la{YukS4} \eeq For $S_3$ triplet scalars
we will consider the following VEV configuration
\beq
\lan \vec{S}_1\ran =\l V_1~, ~0~, ~0\r ~,~~~\lan \vec{S}_2\ran =\l 0~,~
V_2~, ~0\r ~,~~~\lan \vec{A}\ran =\l 0~,~ V~, ~ {\rm i}V\r ~.
\la{S4VEVs}
\eeq
This VEV structure can be obtained from a simple
superpotential. For example, by introducing the singlets superfields
$X_1, X_2, X_A$ with ${\cal U}(1)$ charges $Q_{X_1}=Q_{X_2}=0,
Q_{X_A}=-2q_A$ respectively, allowed superpotential couplings
are\footnote{Similar interactions  within different scenarios have
been considered in refs. \cite{Barbieri:1999km},
\cite{Babu:2007zm}.} \beq W(\vec{S}_{1,2}, \vec{A})=X_1\l
\vec{S}_1^2-V_1^2\r +X_2\l \vec{S}_2^2-V_2^2\r +X_A\vec{A}^2~.
\la{S12Asuppot} \eeq Imposing $F$-flatness conditions one can easily
see that solutions in (\ref{S4VEVs}) are obtained with $\lan
X_{1,2}\ran =\lan X_A\ran =0$. Substituting (\ref{S4VEVs}) in
(\ref{YukS4}) we obtain \beq
\begin{array}{ccc}
 & {\begin{array}{ccc}
\hs{-1.2cm}N_1 ~& N_2
\end{array}}\\ \vspace{0mm}
M_{\nu }=
\begin{array}{c}
l_1\\ l_2\\l_3
\end{array}\!\!\!\!\!\! &{\left(\begin{array}{cc}
\hs{-0.1cm}V_1\ep ~&0
\\
\hs{-0.2cm}V_2\ep ~ &
\hs{-0.1cm}V
\\
\hs{-0.1cm} 0~&{\rm i}V
\end{array} \right)\fr{\ep^n }{M_{\rm Pl}} h_u \! }~,
\end{array}
\la{lNhuS4} \eeq Because of the absence of (1,2) and (3,1) entries
all complex phases can be rotated away from the mass matrix
(\ref{lNhuS4}). Comparing now Eq. (\ref{lNhuS4}) with (\ref{lNhu})
we have $\al =1$ and $b=0$. Therefore, using (\ref{U23}) and
(\ref{13estim}) we get two predictive relations \beq \tan
\te_{23}\simeq 1~, \la{23pred} \eeq and \beq
 \tan \te_{13} \simeq \fr{\De m_{\rm sol}^2}{\De m_{\rm atm}^2}\fr{\tan \te_{12}}{1+\tan^2 \te_{12}} ~.
 \la{13pred}
 \eeq
Therefore, the atmospheric mixing  is maximal ($\te_{23}\simeq
\pi/4$) which is  favored by the neutrino data. For the third mixing
angle we obtain \beq \te_{13}=0.015 \cdot \l 1^{+0.53}_{-0.28}\r ~.
\la{13numPred} \eeq Uncertainty in (\ref{13numPred}) is mostly due
to uncertainties in the measured values of $\De m_{\rm atm}^2$ and
$\sin^2\te_{12}$ in Eq. (\ref{atm-sol-data}). This prediction can be
tested in the planned future neutrino experiments. Also, more
accurate measurements of the solar and atmospheric mass difference
squares, and the solar mixing angle will tighten the range of this
prediction.

\vs{0.1cm}

In conclusion, we have shown that extension of the neutrino sector
with two RHN states and specifically selected flavor symmetries can
provide naturally bi-large neutrino mixings and suppressed value of
$\te_{13}$. The flavor symmetry ${\cal U}(1)\tm S_4$ looks very
promising for obtaining accurate predictions of the $\te_{23}$ and
$\te_{13}$ angles. We note that it is important that the possible
corrections from the charged lepton sector maintain the predictions
derived in the neutral lepton sector. This can be insured by
specific breaking of the flavor symmetry in the charged sector (see
for instance \cite{Babu:2007zm}). It would be interesting to see how
the ideas suggested in this paper might work within various Grand
Unified Theories.

\vs{0.5cm}

\hs{-0.7cm}{\bf Acknowledgments}

\vs{0.2cm} \hs{-0.7cm} The work is supported in part by DOE grants
DE-FG002-04ER41306 and DE-FG02-04ER46140.


\bibliographystyle{unsrt}

\end{document}